# Very-low-field MRI scanners: from the ideal to the real permanent magnet array

U. Zanovello, A. Arduino, V. Basso, L. Zilberti, A. Sola, A. Agosto, L. Toso, O. Bottauscio

***Abstract*— Very-low-field MRIs are becoming increasingly popular due to their portability and adaptability to different environments. They are being successfully used for various clinical applications, leading to a paradigm shift in the way imaging care is typically performed. The development of low-cost MRI scanner prototypes began a few years ago, with some interesting and promising open-source projects emerging in both hardware and software design. Using permanent magnets (PMs) to generate the static magnetic field $B_0$ can substantially reduce the manufacturing cost while achieving satisfactory homogeneity.**

**This article aims to explore the reasons behind discrepancies between magnet design and prototype performance in terms of magnetic field homogeneity. Understanding the impact of the practical implementation of magnet design could inform the development of more tolerant designs in future, simplifying subsequent $B_0$ shimming procedures or even making them unnecessary.**

**This work also evidences the impact of using different numerical model approximations in the modelling phase, proving how they also impact the quality of the design outcomes.**

## I. INTRODUCTION

MAGNETIC resonance imaging (MRI) technology is developing in two different directions. Ultra-high-field MRI scanners allow for enhanced image resolution, improved signal-to-noise ratio (SNR) and better contrast than conventional MRI scanners. Conversely, “low-field” MRI, whose technological evolution is summarized in [1], are becoming increasingly popular due to their portability, affordability, safety and adaptability to different environments [1,2]. The use of artificial intelligence (AI) for image reconstruction, combined with properly characterized hardware, has contributed to the growing interest in these solutions, as evidenced by the extensive literature on the subject published in recent years [2–8]. The attention paid to “low-field” MRI led to the organization of the first ISMRM Workshop on Low-Field MRI in 2022 [9].

“Low-field” MRI are conventionally classified [2, 10] as low-field (LF, < 0.5 T), very-low-field (VLF, 10÷100 mT) and ultra-low-field (ULF, < 10 mT) scanners.

Recently, “low-field” scanners have been successfully employed in various clinical applications, including the diagnosis of brain tumors and stroke [11], neuroimaging [12], the evaluation of Alzheimer's disease [13] and dental imaging [14]. Moreover, their portability introduces a point-of-care model, leading to a paradigm shift in the way imaging care is typically performed [15-19].

Despite being more affordable than their high-field counterparts, proprietary “low-field” scanners are still relatively expensive, limiting their use in low- and middle-income countries. The development of low-cost MRI scanner prototypes began a few years ago with some interesting and promising open-source projects involving both hardware and software design. Recent results have shown image quality comparable to that of commercially available scanners [20-22].

Although adopting permanent magnets (PMs) to generate the static magnetic field ($B_0$) can substantially reduce the manufacturing cost [23–26], achieving satisfactory magnetic field homogeneity can be challenging due to the finite diameter-to-length ratio of Halbach arrays and the use of a limited number of PMs. Field inhomogeneity decreases the SNR and causes image distortion when using fast Fourier transform (FFT)-based reconstructions, thus affecting overall image quality [27, 28].

Design optimizations [23, 29-33] and flexible passive/active shimming approaches [34] are paramount in getting field homogeneity. A recent review [35] provides a compilation of relevant concepts in designing Halbach multipoles for magnetic resonance applications.

Besides achieving satisfactory magnetic field homogeneity, proper design of the different MR hardware proves to be essential for producing high-quality images [36]. Among them, for example, the design of the radiofrequency coil to compensate for the low signal-to-noise ratio (SNR) [37], the design of the gradient coils to minimize distortion caused by ferromagnetic structures [38], and the use of acquisition techniques based on AI tools [39].

Proper metrological characterization of the scanner components is a valuable step in hardware development, proving fundamental information in solving potential issues during construction and testing the numerical tools used for system design. The latter is particularly critical when it comes to the main magnet. Manufacturing tolerances and the intrinsic magnetic variability of the PMs, together with their dependence on temperature, can in fact be detrimental to scanner performance.

The project 22HLT02 A4IM has received funding from the European Partnership on Metrology, co-financed by the European Union's Horizon Europe Research and Innovation Programme and by the Participating States. (*Corresponding author:* O. Bottauscio).

U. Zanovello, A. Arduino, V. Basso, L. Zilberti, A. Sola, A. Agosto, L. Toso and O. Bottauscio are with the Istituto Nazionale di Ricerca Metrologica, Torino, 10135, Italy (e-mail: o.bottauscio@inrim.it).

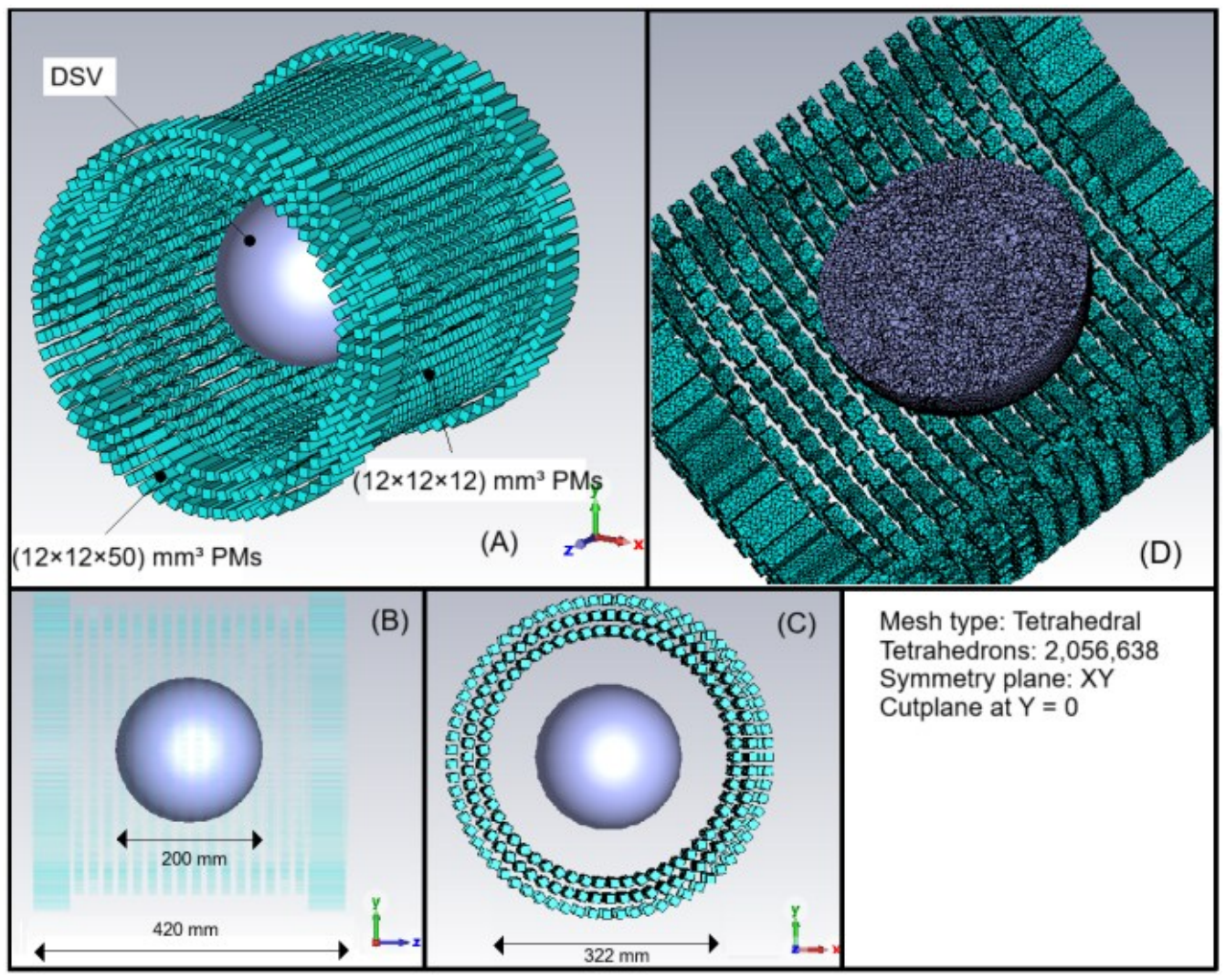

**Fig. 1.** Magnet array configuration with the considered spherical DSV. (A) 3D view, (B) lateral view, with total magnet length and DSV diameter size, (C) front view with the bore internal diameter, (D) Finite Element mesh adopted in the modeling phase described in Section IV. The magnetic flux density $B_0$ in the DSV is mainly oriented along the $x$-axis.

This article fits this context by focusing on the characterization of magnet performance in terms of magnetic field homogeneity for VLF scanners using PMs. The main objective is to investigate the sensitivity of the magnetic field homogeneity to various factors, exploring the reasons behind discrepancies between performance of the reference design and the built prototype. To this scope, the analysis has taken as a reference the VLF magnet designed in the framework of the Open-Source Imaging Initiative (OSI²) [40].

Understanding the impact of the choices made during the construction of the magnet can lead to more effective designs. Furthermore, improving the agreement between actual and expected behaviors of the magnet would simplify the subsequent $B_0$ shimming procedure, for example by reducing the number and size of shimming magnets and the number of shimming iterations, or even making the shimming phase unnecessary. In this regard, even if the results of this study refer to specific VLF magnet implementation, they prove to be relevant and useful to the wider portable MRI community.

An additional aim of this study is to quantify the impact of using different numerical model approximations in the design phase, revisiting concepts that are most frequently used in the design of electrical machinery [41]. Specifically, while some assumptions may simplify the numerical model and improve its performance in terms of computational time, this paper demonstrates how they also impact the quality of the obtained results.

## II. Magnet array

### A. Magnet array layout

The OSI² v2.1 magnet array called 'ROMA' (Rotation-Optimized Magnet) [42], was considered for this study. The design of the 'ROMA' array was optimized by Leiden University Medical Center and the mechanical frame has been engineered by the Physikalisch-Technische Bundesanstalt (PTB) in Berlin. To the purpose of the analysis here reported, this magnet has been replicated and characterized at INRiM to compare experimental and simulation outcomes.

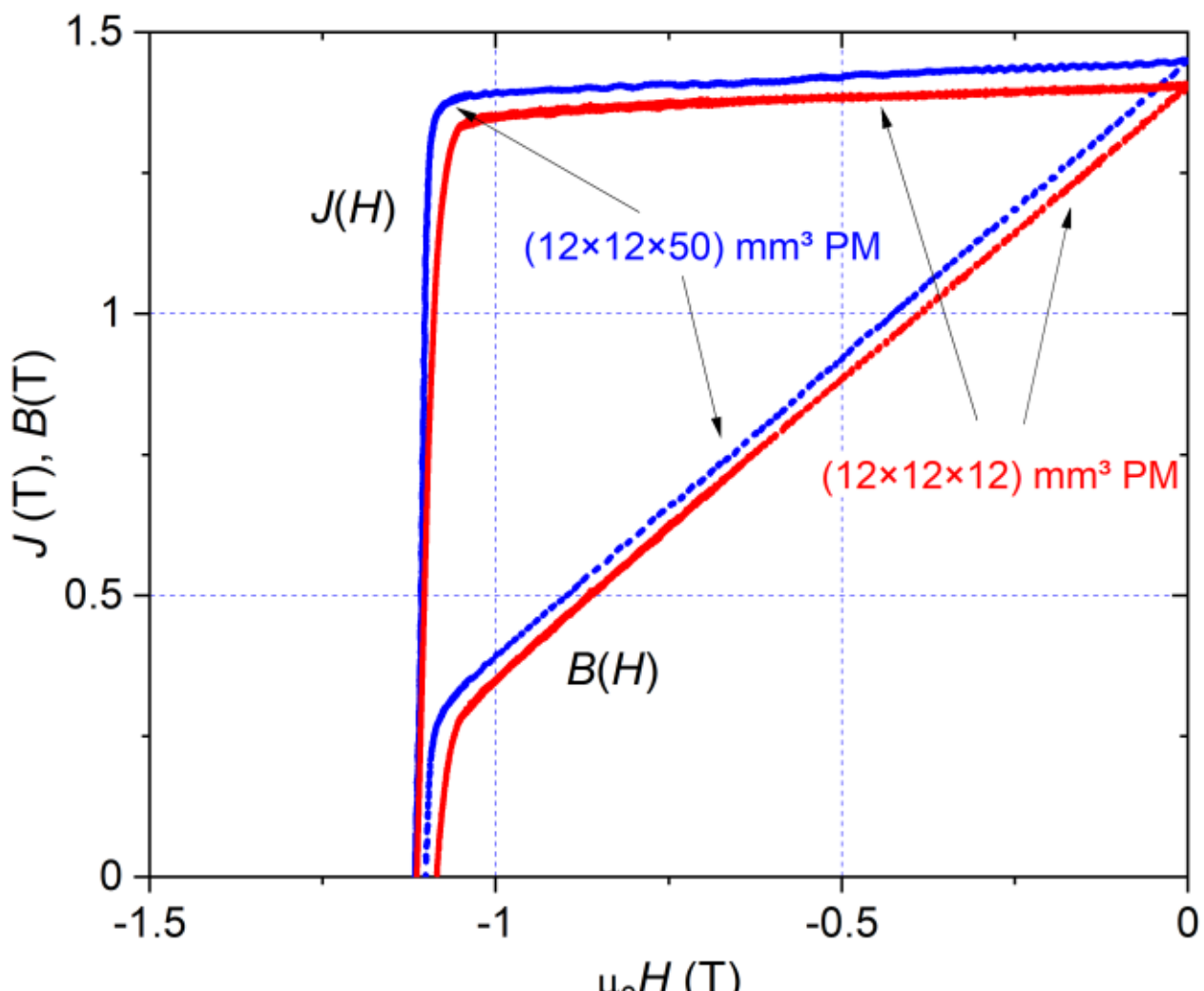

**Fig. 2.** Magnetic characteristics of the two samples of NdFeB N52 PMs. Curves reported in the figures refer to 24 °C.

In the 'ROMA' magnet configuration, the orientation of the individual PMs within the array does not adhere to the Halbach theory or its Mandhala discretization [43]. Instead, the orientation of the PMs is the result of an optimization procedure designed to maximize the magnetic field homogeneity within a 200 mm DSV (Diameter of Spherical Volume) at the center of the magnet.

The ROMA array comprises 1936 NdFeB N52 PMs measuring (12×12×12) mm³, arranged in two circular layers along 16 rings. The rings are made of polypropylene (PP) using a computer numerical control (CNC) milling process.

Two additional rings close the array at the extremities. These rings contain 384 NdFeB N52 PMs measuring (12×12×50) mm³ arranged on three circular layers. Figure 1 shows an overview of the entire setup. The bore diameter is 322 mm, with an axial length of 420 mm.

The mass of the magnet array is approximately 100 kg. The target magnetic flux density ($B_0$) at the isocenter is approximately 50 mT, oriented along the $x$-axis.

### B. PMs magnetic characteristics

The magnetic characteristics of the NdFeB PMs used to build the array were measured using the hysteresisgraph method at temperatures ranging from 18 °C to 50 °C.

The following relationships fit the dependence of the residual polarization $J_r$ and magnetic coercivity $H_c$ on temperature $T$.

$$J_r = J_{r0}\left[1 + K_J\,(T - T_{\text{ref}})\right] \quad (1a)$$

$$H_c = H_{c0}\left[1 + K_{H1}\,(T - T_{\text{ref}}) + K_{H2}\,(T - T_{\text{ref}})^2\right] \quad (1b)$$

with $J_{r0}$ ($H_{c0}$) equal to 1.431 T (–955.3 kA/m) and 1.462 T

(−941.2 kA/m) for the (12×12×12) mm³ and (50×12×12) mm³ PMs, respectively. $T_{ref}$ is the reference temperature at which $J_{r0}$ and $H_{c0}$ were measured ($T_{ref}$ = 18 °C in our case), $K_J$ = −1.26 $10^{-3}$ °C$^{-1}$, $K_{H1}$ = −0.01 °C$^{-1}$ and $K_{H2}$ = +3.8 $10^{-5}$ °C$^{-2}$.

Figure 2 shows the magnetization curves of (12×12×12) mm³ and (12×12×50) mm³ PM samples related to 24 °C. Magnetization curves are shown in terms of magnetic flux density ($B$) and magnetic polarization ($J$) versus magnetic field ($H$). Despite being provided by the same producer, the PMs exhibit some differences in their magnetic characteristics.

## III. Experimental characterization of the magnet array

The magnet array was assembled in accordance with the guidelines provided in [44]. Each PM was tested before being inserted in the rings using a properly designed magnet test station [45, 46]. The assembled array was placed in a 3D automatic movement system, with the *z*-axis aligned with the vertical axis of the plotter to perform magnetic field measurements in the DSV. Figure 3 shows a photograph of the array inside the 3D movement system.

Measurements were performed using a SENIS 3MH6-E teslameter, which has 3-axis Hall probes packed in a volume of (100×10×100) µm³ to ensure optimum spatial resolution. The probe was fixed to the movement system using *ad hoc* support, and a Python program synchronized the movement and acquisition processes. The acquisition pipeline allowed sufficient idle time between consecutive measurements to ensure that probe oscillations due to probe motion were fully damped. Before the acquisition, the magnet array was left in the laboratory for over 24 hours. This allowed all the PMs to reach the laboratory temperature of 24 °C ± 2 °C. Additionally, the Python software recorded the temperature measured by the sensor integrated within the magnetic field probe for each magnetic field acquisition, resulting in an average temperature of 23.7 °C ± 0.3 °C. This information was used in both simulation and calculation of the uncertainty budget.

The magnetic field was measured inside the 200 mm DSV in uniformly distributed 4224 points, defined by an equally spaced cartesian grid with steps equal to 10 mm along the *x*, *y* and *z* directions. The grid resolution was chosen as a trade-off between sufficient spatial resolution and elapsed measurement time, considering that the entire acquisition took approximately 10 hours to complete. The contribution of the Earth's magnetic field was removed from the measured data.

The field homogeneity of $B_x$ in the DSV was quantified by the following two metrics:

$$DIS_1 = \frac{Max(B_x) - Min(B_x)}{Mean(B_x)} \quad (2a)$$

$$DIS_2 = \frac{Std(B_x)}{Mean(B_x)} \quad (2b)$$

All the statistics (*Max*, *Min*, *Mean* and *Std*) were evaluated on the measured sampled values. The use of a uniform grid enable us to compute the mean and standard deviation of the spatial distribution of the magnetic field component $B_x$ in the DSV directly as sample mean and sample standard deviation of the measured values.

The $DIS_1$ parameter is widely adopted by the MRI community [14, 20, 33, 43] and facilitates comparison with other implementations. Whereas it shows correlations with image quality in terms of signal intensity and distortions [27, 28], it is based on local values of the $B_0$ field (maximum and minimum values). The MRI signal is the result of all the signals emitted by the excited magnetization vectors. This consideration justifies the adoption of the integral $DIS_2$ parameter, which incorporates the effects of all spatial values, as an alternative to $DIS_1$.

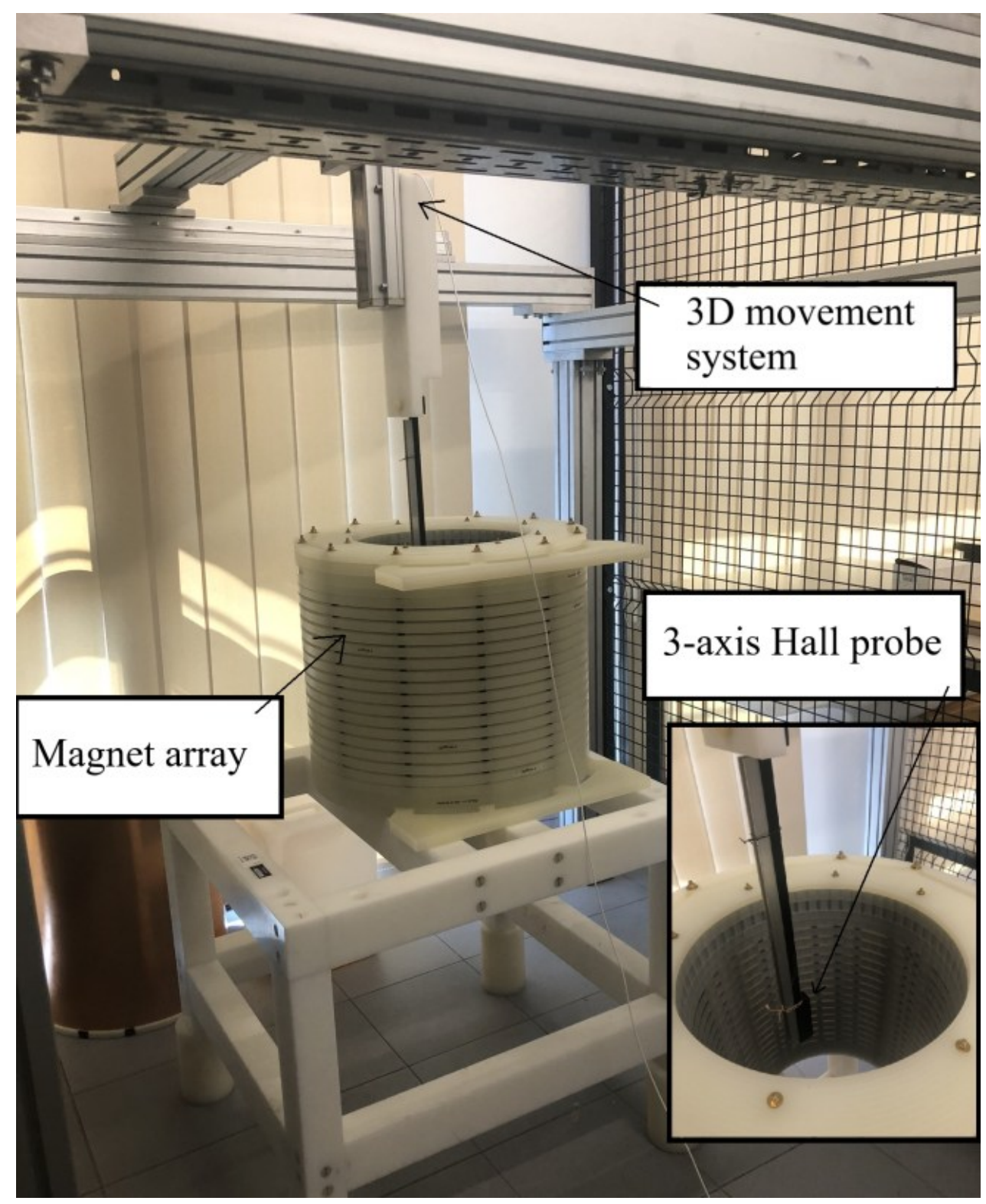

**Fig. 3.** Photograph of the assembled array placed within the 3D movement system used to perform the magnetic measurements in the DSV.

It is worth noting that $DIS_2$ is controlled by $DIS_1$ in the sense that $DIS_2 \leq DIS_1$. In particular, the equality is only verified when $DIS_1 = 0$. Consequently, it is possible to have a small $DIS_2$ and a large $DIS_1$, but not vice versa.

The average magnetic flux density measured was 48.056 mT, with a relative uncertainty of 166 ppm. The experimental $DIS_1$ and $DIS_2$ values were 24600 ppm ± 1300 ppm and 4400 ppm ± 1300 ppm, respectively. The expanded uncertainty values (95 % level of confidence) were estimated according to the procedure described in Appendix A. The measured $DIS_1$ value is comparable with the one obtained in other practical realization of scanners before shimming (see for example [7, 20]).

## IV. Numerical models

In the simulation environment, the *z*-axis is oriented along the bore axis, with the main component of the static magnetic field oriented along the *x*-axis (see Figure 1). The magnetic field homogeneity was theoretically evaluated in the 200 mm DSV by adopting two different modelling approaches: the finite element method (FEM) and the dipole model. These approaches were further specialized in the way they dealt with the nonlinear magnetic characteristics of the PMs. The FEM approach is more computationally demanding but provides a more reliable solution for the actual PM geometry. The dipole model enables faster calculations by using closed-form approximations to compute the magnetic field generated by the PMs.

*A. Finite element solver*

Due to its better geometrical accuracy, the FEM solution was used as a reference for validating the dipole models. The geometric configuration of the magnet array was imported into CST Studio Suite [47]. A bounding box with sides of 1 m was set around the magnet array with open boundary conditions. A symmetry constraint was imposed on the *xy*-plane at $z = 0$, and the structure was then discretized into conformal tetrahedral elements. A magnetostatic solver with mesh adaptation was then used. The solver accuracy was set to $10^{-9}$, resulting in 2.1 $10^6$ tetrahedrons after mesh refinement. The average mesh quality (ratio between inradius and circumradius) was 0.71, with a minimum and maximum edge lengths of 0.78 mm (close to the PMs) and 100 mm (towards the external boundaries). Figure 1D provides a view of the mesh.

The solver modelled the PM characteristics with two different levels of approximation:

- linear magnetic characteristics extrapolated from the data in Fig. 2 around the remanence point, as

$$B = \mu_0 H + J = \mu_0 H + \mu_0 \mu_M H + J_r \tag{3}$$

where $\mu_M$ is equal to 0.0223 and 0.0168, for the (12×12×12) mm³ and (50×12×12) mm³ PMs, respectively, $\mu_0$ is the magnetic permeability of vacuum and $J_r$ was assumed to be equal to the value obtained by Eqn. (1a) at the working temperature.

- nonlinear behavior deduced from the measurements.

*B. Dipole model*

The dipole solver is based on the work of Engel-Herbert and Hesjedal [48], in which the authors propose an analytical solution for the magnetic field produced by a bar-shaped permanent magnet under the assumption of uniform magnetization. A local Cartesian coordinate system ($u$, $v$, $w$) with associated unit vectors ($\boldsymbol{u}$, $\boldsymbol{v}$, $\boldsymbol{w}$) is centered on the barycenter of the magnet bar, having dimensions ($2L_u$, $2L_v$, $2L_w$). Being the local $v$-axis oriented along the magnetization $J_v$, the analytical expressions for the magnetic field components $H$ at a generic point ($u$, $v$, $w$) are given by:

$$H_u(u,v,w) = \mu_0 \frac{J_v}{4\pi} \sum_{k,l,m=1}^{2} (-1)^{k+l+m} log\left(w + (-1)^m L_w + \sqrt{[u + (-1)^k L_u]^2 + [v + (-1)^l L_v]^2 + [w + (-1)^m L_w]^2}\right) \tag{4a}$$

$$H_v(u,v,w) = -\mu_0 \frac{J_v}{4\pi} \sum_{k,l,m=1}^{2} (-1)^{k+l+m} \frac{[v+(-1)^l L_v]\cdot[u+(-1)^k L_u]}{|v+(-1)^l L_v|\cdot|u+(-1)^k L_u|} \times arctan\left(\frac{|u+(-1)^k L_u|\cdot[w+(-1)^m L_w]}{|v+(-1)^l L_v|\cdot\sqrt{[u+(-1)^k L_u]^2+[v+(-1)^l L_v]^2+[w+(-1)^m L_w]^2}}\right) \tag{4b}$$

$$H_w(u,v,w) = \mu_0 \frac{J_v}{4\pi} \sum_{k,l,m=1}^{2} (-1)^{k+l+m} log\left(u + (-1)^k L_u + \sqrt{[u + (-1)^k L_u]^2 + [v + (-1)^l L_v]^2 + [w + (-1)^m L_w]^2}\right) \tag{4c}$$

The above equations represent the 'external' magnetic field generated by a given permanent magnet within the surrounding volume. They can be summarized in vector form as follows:

$$\boldsymbol{H}_{\mathrm{ext}} = \begin{pmatrix} H_{u,\mathrm{ext}} \\ H_{v,\mathrm{ext}} \\ H_{w,\mathrm{ext}} \end{pmatrix} = J_v \begin{pmatrix} f_u \\ f_v \\ f_w \end{pmatrix} = J_v \, \boldsymbol{F} \tag{5}$$

$\boldsymbol{F}$ is a column array whose elements are derived from equations (4).

Similarly to FEM, the dipole model was extended to account for the interaction between the surrounding magnets, modelling both the linear and nonlinear characteristics of the NdFeB N52 magnetization curve. Details about the dipole model extensions are reported in Appendix B, together with a simple diagram showing how FEM and dipole models approximate the PMs array.

*C. Dipole approach vs. finite element results*

The dipole approach was validated against FEM considering the data of the ROMA array configuration published in [42]. To comply with the assumptions of the optimization procedure that

TABLE I

MEASURED AND SIMULATED VALUES OF THE MEAN VALUE OF $B_X$ IN THE DSV AND FIELD HOMOGENEITY WITH THE TWO ADOPTED METRICS ($DIS_1$ AND $DIS_2$). THE VALUES CORRESPOND TO THE PM CHARACTERISTIC AT 23.7 °C. IN THE EXPERIMENTS, THE EXPANDED UNCERTAINTY (WITH 95 % LEVEL OF CONFIDENCE), ESTIMATED AS DESCRIBED IN APPENDIX A, IS REPORTED. BOTH FEM AND DIPOLE SIMULATIONS ARE REPORTED.

| Case | *Mean*($B_x$) (mT) | $DIS_1$ (ppm) | $DIS_2$ (ppm) | Total computational time (*) (hh:mm:ss) |
|---|---|---|---|---|
| Experiments | 48.056 ± 0.008 | 24600 ± 1300 | 4400 ± 1300 | - |
| Ideal Dipole | 48.938 | 266 | 26.4 | 00:00:15 |
| Dipole with reaction linear *H-J* curve | 48.581 | 1180 | 209 | 00:00:29 |
| FEM with linear *H-J* curve | 48.205 | 1246 | 184 | 00:34:00 |
| Dipole with reaction nonlinear *H-J* curve | 47.908 | 1720 | 316 | 00:01:07 |
| FEM with nonlinear *H-J* curve | 47.905 | 2165 | 394 | 01:24:00 |

*Note: (*) They include all computational phases (pre-processing, solver and post-processing).*

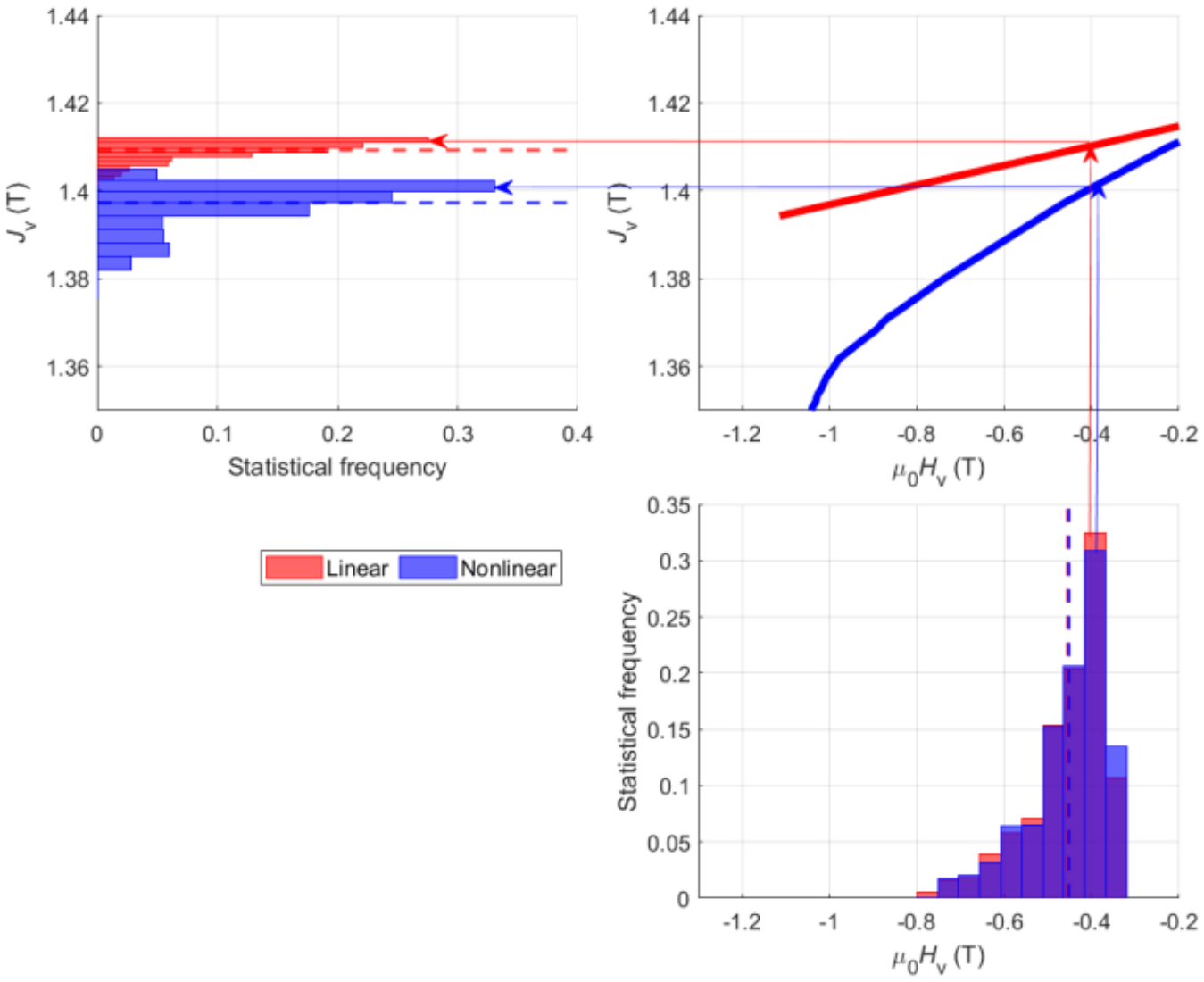

**Fig. 4.** Statistical distribution of the working point ($H_v$, $J_v$) of each single PM using linear and nonlinear PM characteristics. Upper left plot: statistical frequency of $J_v$ values, Upper right plot: *H-J* characteristic of the PMs, Bottom right plot: statistical frequency of $H_v$ values. The dashed lines are the mean values of $J_v$ and $H_v$. The arrows illustrate the correspondence between $H_v$ and $J_v$ values within a single magnet for the linear and nonlinear cases.

led to the magnet design, all simulations assumed the same magnetic characteristics for all the permanent magnets (PMs) equal to those of the (12×12×12) mm³ PMs. The results were compared in terms of the relative discrepancy in the L2-norm of the *x*-component of the magnetic flux density ($B_x$), sampled in the same *N* points ($N = 4224$) mapped in the experiments. Simulations were performed with PMs characteristic at 23.7 °C allowing comparison with experiments.

$$L2-norm = \sum_i^N \left(B_{x,dipole}^{(i)} - B_{x,CST}^{(i)}\right)^2 / \sum_i^N \left(B_{x,CST}^{(i)}\right)^2 \quad (4)$$

In Eq. (4) $B_{x,dipole}^{(i)}$ and $B_{x,CST}^{(i)}$ are the values computed by the dipole model and CST, respectively, using either the linear or nonlinear solver. The relative discrepancies are 4.67 $10^{-7}$ and 3.22 $10^{-6}$ for the linear and nonlinear cases, respectively, indicating good agreement between the two modelling approaches.

Table I summarizes experimental and computed results with increasing level of complexity, starting from the ideal dipole approximation, *i.e.* when the dipoles are assumed to have a working point corresponding to a magnetic polarization equal to $J_r$. This latter approach produces results that are inconsistent with those obtained using FEM.

The differences between the results obtained by assuming linear and nonlinear PM characteristics can be explained by the different working points of each PM, which result from their interaction. The plots in Figure 4 show the distribution of the magnetic field ($H_v$) and polarization ($J_v$) components in each PM for the two cases. While the $H_v$ distributions are similar, the $J_v$ distributions differ significantly, with lower mean value for the nonlinear case.

Table I also summarizes the computational times required for the different modelling approaches. The FEM results were obtained using the magnetostatic CST solver (CST Studio Suite 2025), while the dipole results were obtained using a custom-made Matlab R2022a script [49]. The computational times refer to the solutions obtained using an Intel® Xeon® Gold 6430 processor with a frequency of 2.10 GHz and 512 GB of RAM.

With exception of the FEM model, the computational time scales almost linearly with the number of computational points in the DSV. For FEM the computational time is related to the mesh size of the entire domain. In any case, the burden on the measurement process definitely overcomes the one of the computations.

Based on these outcomes, the dipole model was found to be accurate enough for all subsequent analyses. It also considerably reduced the computational burden, enabling Monte Carlo analyses that would not have been feasible with FEM.

## V. Deviations between the Prototype and Reference Design

Despite the great care taken in manufacturing the magnet components and assembling the final array, the comparison between measurements and simulations revealed significant discrepancies in terms of field homogeneity. This section investigates several possible deviations from the ideal structure and their impact on field homogeneity.

We drew up an inventory of the main reasons for deviation based on building, design and manufacturing experience. They can be either deterministic or stochastic in nature. Both cause the array to behave differently from the reference design but, whereas the former can be accurately determined and possibly corrected in a new design, the latter can only be compensated for through shimming strategies.

In the following sections these two sources of deviation are investigated in detail. Equations (1) were applied to perform simulations at an average temperature of 23.7 °C, as measured during the experiments.

### *A. Deterministic causes of deviation*

Three possible deterministic causes of deviation from design to prototypes have been identified: a) the different magnetic characteristics of the two used PMs types, b) the ring positioning along *z*-axis, c) the modification of PM orientations due to mechanical forces. These three causes were studied in strict sequence, with each subsequent effect evaluated on top of the previous one.

All of them have a deterministic character, because they can be quantified and, in principle, compensated during the design phase.

#### 1) **Different PM characteristics**

The hysteresisgraph measurements in Section II.B revealed different behavior between the (12×12×12) mm³ and (12×12×50) mm³ PMs, even though they are both N52 type and supplied by the same manufacturer. This significant difference in behavior affects the magnetic field homogeneity due to the unbalanced contribution of the short and long PMs. Considering the different PM characteristics, the values of $DIS_1$ and $DIS_2$ increased from 1720 ppm to 6960 ppm and from 316 ppm to 1340 ppm, respectively.

#### 2) Ring positioning along *z*-axis

Another cause of deviation from ideal behavior that was investigated is the position of the rings along the *z*-axis. Indeed, the strong attraction or repulsion between the adjacent rings slightly affects their distance, which could not be compensated for using the threaded rods to pack the frames. Therefore, the position of each ring was measured using a caliber on the built array. The small deviation of the central position of each ring from the reference design was determined in order to repeat the computation. This source of deviation has an opposite effect on $DIS_1$ and $DIS_2$ values, leading to a small reduction of the values obtained with the previous correction from 6960 ppm to 6520 ppm and from 1340 ppm to 1150 ppm, respectively.

#### 3) Rotation of PM orientation due to mechanical torque

The small rotations of the PMs with respect to their reference orientation, which are caused by their mechanical interactions, were investigated. These rotations can be estimated by computing the torque acting on each PM, because of the interactions with all other PMs. Figure 5 shows the behavior of the torque acting in the *xy*-plane on each single PM. In the figure only a selection of rings is shown for brevity. The other torque components, acting on *xz*- and *yz*-planes, are lower (always less than 10 %). Therefore, assuming that the mechanical constraints provided by the frame containing the magnets are similar along the three axes of rotation, the effects of such torques are neglected. This choice, which has been subjected to verification a posteriori, is also justified by the fact that the magnetization of the magnets is orthogonal to the *z*-axis, making rotations in the *xy* plane more relevant for the correct determination of the field in the DSV.

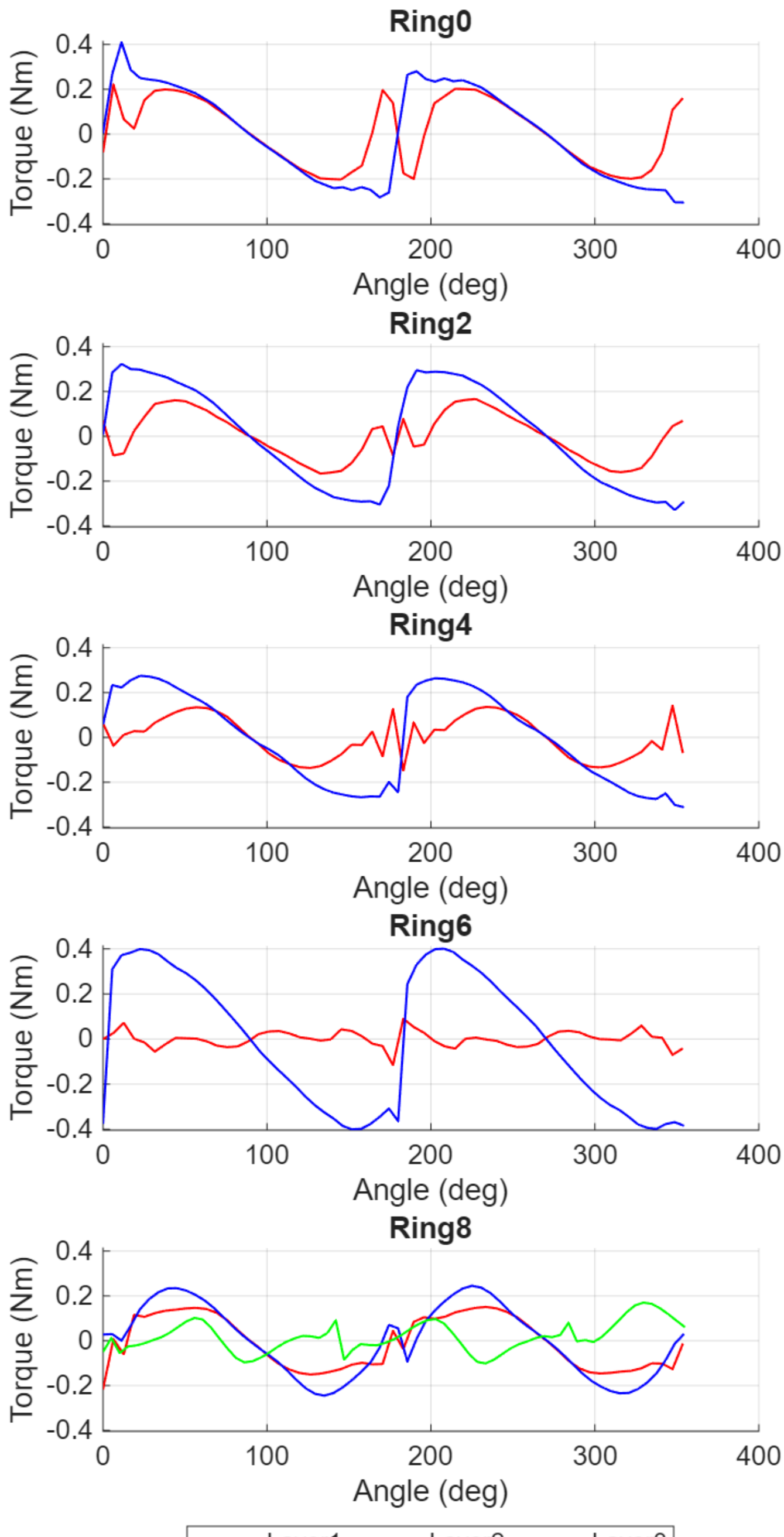

**Fig. 5.** Behavior of the torque acting in the *xy*-plane on each PM due to the interaction with all other PMs. The figure shows the torque acting on the even rings placed on the positive *z*-axis (from the one closest to the isocenter, Ring0, to the one at the extremity, Ring8). Lines connecting scattered data are just a guide for the eyes.

PM rotation is a consequence of the backlash within the frame pocket. Its magnitude depends on several factors, some of which are stochastic, such as manufacturing tolerances and variability in PM dimensions, while others depend on design and mechanical requirements. In the realized array, the magnet pockets provided a good fit for the PMs while still allowing easy manual insertion. Due to other difficult-to-account-for influencing factors, deciding upon a reference rotation angle is not an easy task. For example, PP manufacturing is likely to increase CNC tolerance, and the force applied by the PMs to the frame may deform the pockets slightly. Therefore, the investigation considered three values of reasonable rotation angles of the PMs inside their pockets: ±0.8°, ±1.0° and ±1.2°. These angles correspond to pockets that are 0.13 mm, 0.17 mm and 0.2 mm larger than the 12 mm side of the PM, respectively.

The amplitude of the rotational angles is applied identically to all PMs without regard for the torque acting on each individual PM. However, the sign of the rotation with respect to the reference orientation is correlated with the sign of the torque acting on each specific PM.

Figure 6 illustrates the impact of the PM rotations. The plots on the left show the values of $B_x$ sampled on the 4224 measurement points within the DSV. Measured outcomes can be compared with the computed values obtained with the theoretical orientation of PMs and the ones obtained when PMs are rotated of ±0.8° or ±1.2°, according to the sign of the computed torque. These results are obtained having already included the different PM characteristics and adjustment of *z*-position of the rings as discussed in the previous sub-sections.

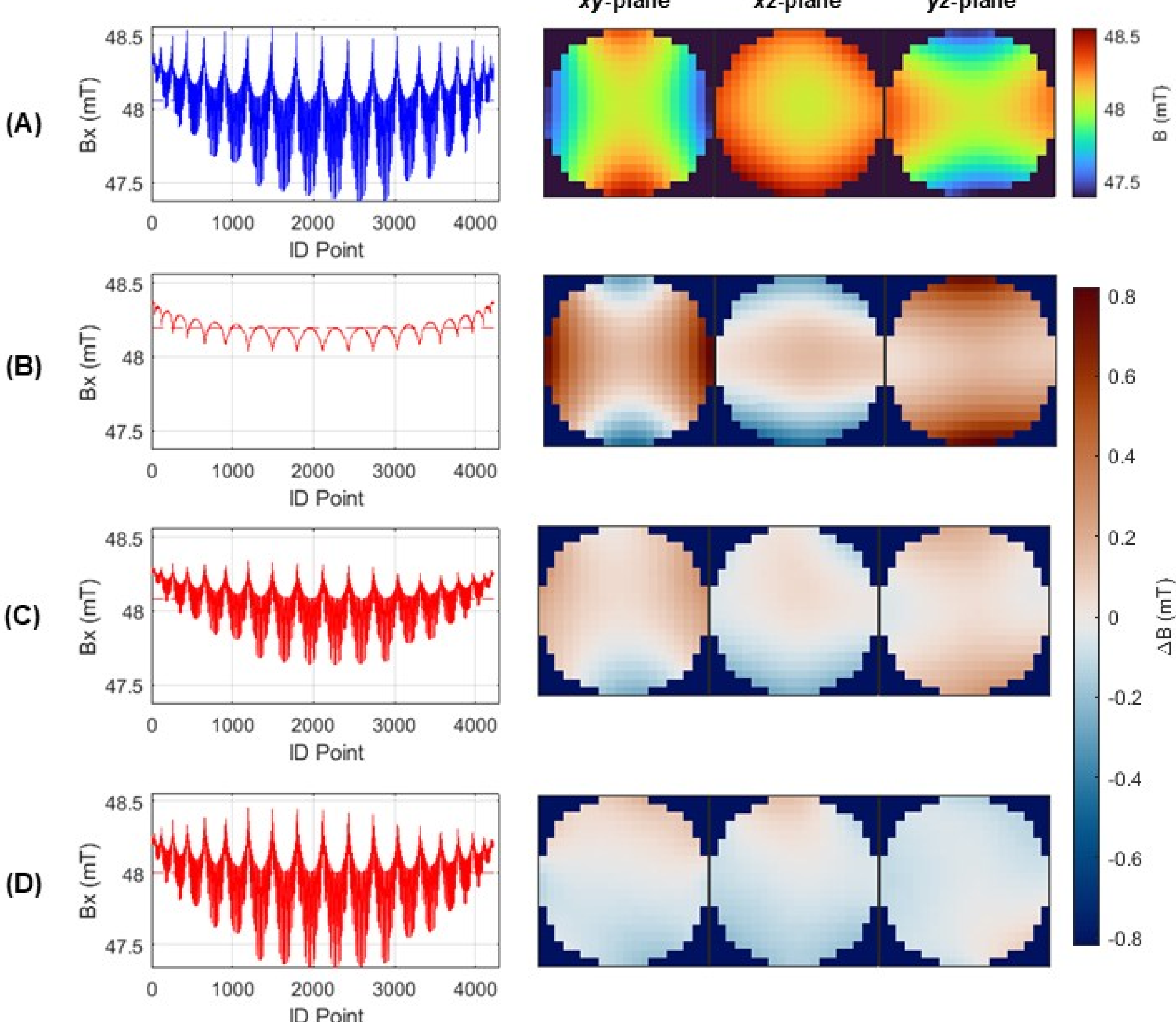

**Fig. 6.** Comparison between measured and computed $B_x$ values within the DSV. On the left side the measured and computed values sampled on the 4224 measurement points. On the right side the corresponding maps in the *xy*-, *xz*- and *yz*-planes. From top to bottom row: (A) measured sampled values and corresponding maps of $B_x$; from (B) to (D) computed sampled values of $B_x$ and corresponding maps of the difference $\Delta B$ with respect to measurements, with PMs having the theoretical orientation (B), with PMs rotated by ±0.8° (C), with PMs rotated by ±1.2° (D).

We notice that the good agreement between the $B_x$ trends shown in figures 6A and 6D provides, retrospectively, confirmation of the validity of the choice to consider only the torque components that act in the *xy*-plane.

In the same figure, the impact on the spatial distribution of $B_x$ on the *xy*-, *xz*- and *yz*-planes is also shown. Adjusting the PM rotation, the values of the difference $\Delta B$ with respect to the measurements reduces as proved by the corresponding maps.

Considering all the above-mentioned deterministic cause of variabilities, the values of $DIS_1$ (*resp.* $DIS_2$) were found to be 14670 ppm (*resp.* 2670 ppm) for ±0.8°, 18930 ppm (*resp.* 3340 ppm) for ±1.0° and 23200 ppm (*resp.* 4028 ppm) for ±1.2°. Despite these being rather small angles, the results reveal the significant impact of this effect on the assembled magnet array. This effect also leads to a good agreement between the measured and simulated $B_x$ field along the Cartesian axes (Figure 7) and along *z*-lines outside the DSV (Figure 8) obtained with the rotational angle of ±1.2°.

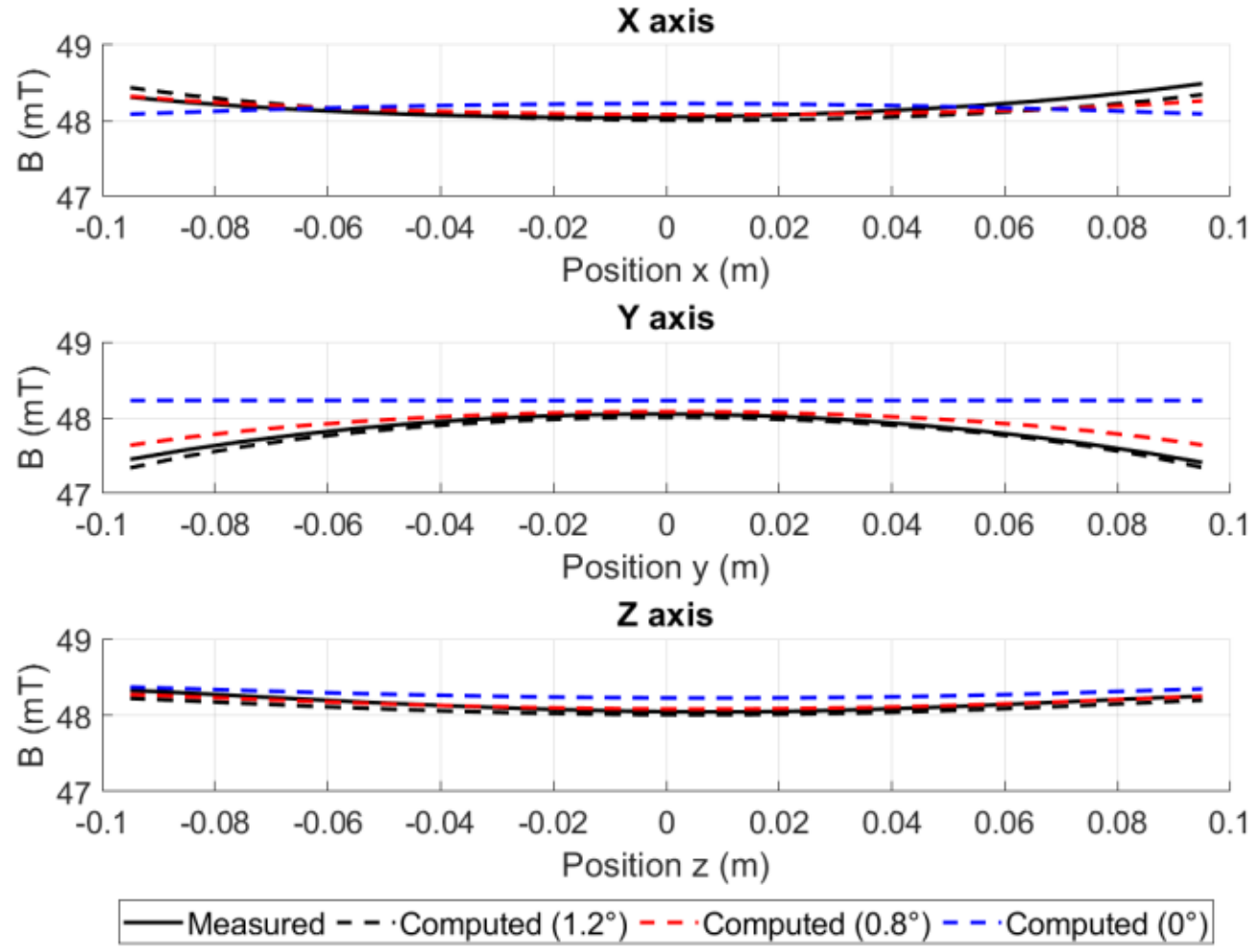

**Fig. 7.** $B_x$ values along the Cartesian axes in the DSV. The measured values (solid lines) are compared with the computed ones with PMs having the ideal orientation (dashed blue lines), with PMs rotated by ±0.8° (dashed red lines), and with PMs rotated by ±1.2° (dashed black lines).

*B. Stochastic causes of deviation*

After including all the deterministic causes of deviation, the following additional stochastic effects were accounted for in the investigations:

a) variability of the residual polarization of each PM sample;

b) measurement uncertainty in the characterization of the PM nonlinear characteristics;

c) uncertainty in each PM rotation;

d) PM positioning due to dimensional tolerances in the CNC machinery used to produce the PP rings.

Unlike the deterministic causes of deviation analyzed in the previous section, these effects cannot be compensated for during the design phase. However, they can be used to assign an uncertainty value to the design prediction.

These effects were all studied separately to quantify their impact, and they were applied on top of the deterministic effects identified in Section V.A, *i.e.* the different PM characteristics, the adjustment of the $z$-position of the rings and the rotational angles of ±1.2° of the PM due to the torque.

The analysis assumed a statistical distribution of the input quantities based on measurements or assumptions. Monte Carlo (MC) simulations were then performed to determine the statistical distribution of the output quantities (i.e. the mean value of $B_x$ in the DSV and the $DIS_1$ and $DIS_2$ values). For each simulation, the input quantity was randomly extracted from its statistical distribution. The MC analysis was based on 1000 extractions. The stability of the MC outcomes versus the number of extractions was verified a posteriori, with the output deviation remaining below 0.3 % when increasing the number of extractions to 10000.

All results are collected in Table II, which shows the expected values expanded uncertainty with level of confidence of 95 % of $Mean(B_x)$, $DIS_1$ and $DIS_2$ in the DSV.

**1) Variability of the residual polarization of PMs**

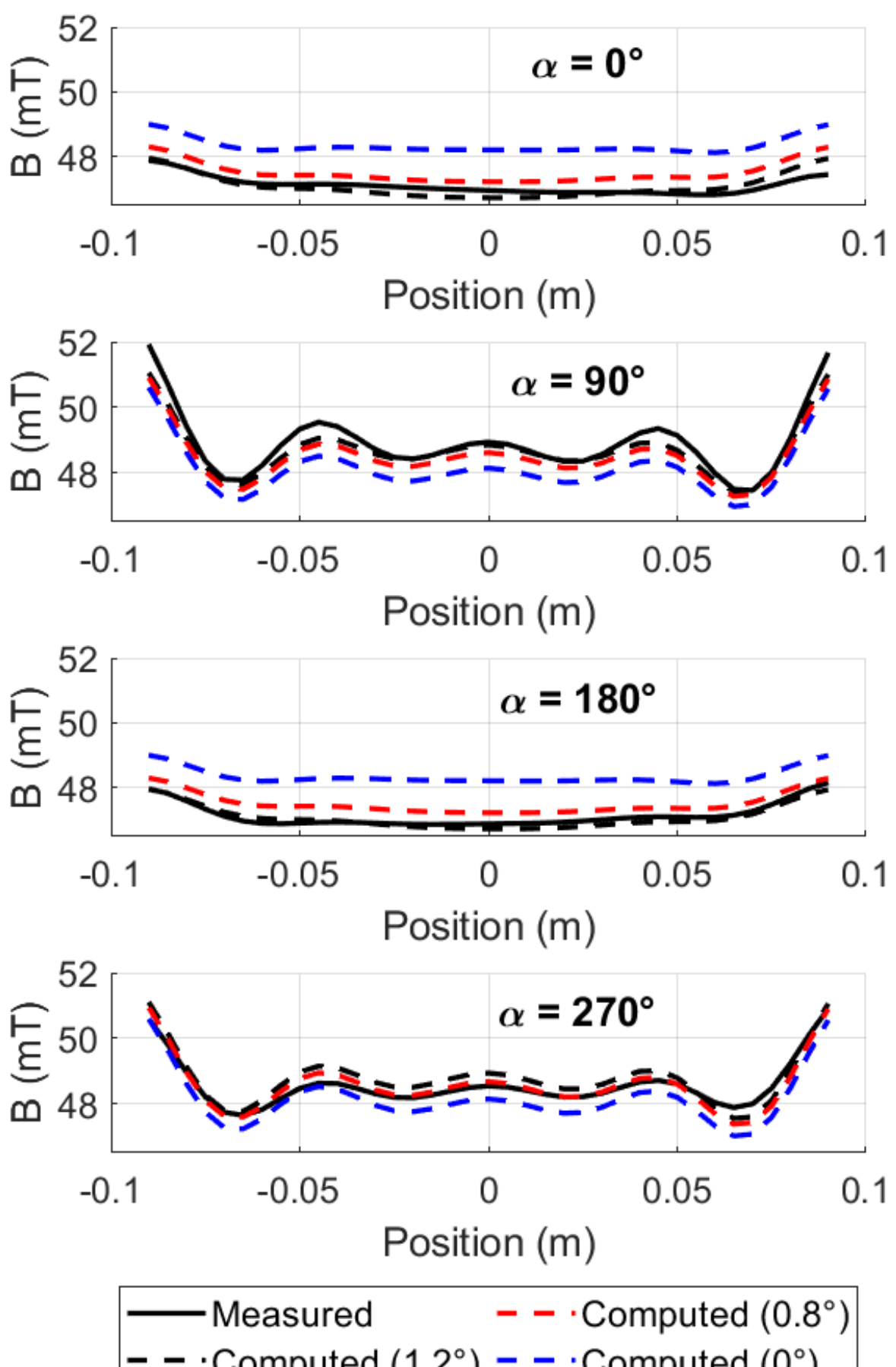

**Fig. 8.** $B_x$ values along four $z$-lines (from -0.09 m to +0.09 m) outside the DSV at a radius equal to 0.12 m and spaced by 90° in the $xy$-plane. For each plot α is the angle (in degrees) with respect to the $x$-axis. The measured values (solid lines) are compared with the computed ones with PMs having the ideal orientation (dashed blue lines), with PMs rotated by ±0.8° (dashed red lines), and with PMs rotated by ±1.2° (dashed black lines).

Although the magnetic remanence of the PMs is a specification provided by the manufacturer and measured by the hysteresisgraph on a selected sample, the actual value of each PM may differ slightly due to manufacturing and magnetization processes, storage conditions, etc. As a hysteresisgraph measurement takes a long time, it was not feasible to perform a complete magnetic characterization of all PMs. Therefore, the dispersion of the magnetic remanence of 20 samples randomly extracted from the (12×12×12) mm³ batch and 10 samples from the (12×12×50) mm³ batch was estimated by characterizing their working point. The magnetic moment was measured using the extraction method [50], and the relative standard deviations of the two distributions were found to be 0.59% and 0.26%, respectively.

Monte Carlo (MC) simulations were performed by randomly extracting the value of $J_r^{(i)}$ for each PM from a Gaussian distribution with a mean value (*resp.*, standard deviation) of 1.421 T (8.2 mT) for the (12×12×12) mm³ PMs and 1.451 T (3.8 mT) for the (12×12×50) mm³ PMs.

TABLE II

RESULTS OF THE STOCHASTIC SIMULATIONS. FOR EACH INPUT VARIABILITY THE TABLE COLLECTS THE EXPECTED VALUES AND EXPANDED UNCERTAINTIES (WITH 95 % LEVEL OF CONFIDENCE) OF MEAN($B_X$), $DIS_1$ AND $DIS_2$.

| Input variability | *Mean*($B_x$) (mT) | $DIS_1$ (ppm) | $DIS_2$ (ppm) |
|---|---|---|---|
| PMs residual polarization | 48.01 ± 0.008 | 23300 ± 878 | 4030 ± 92 |
| Measurement uncertainty in PM magnetic characterization | 48.01 ± 0.58 | 23400 ± 1726 | 4080 ± 221 |
| Uncertainty in PM orientation | 48.00 ± 0.010 | 23370 ± 757 | 4032 ± 90 |
| Uncertainty in PM positioning | 48.00 ± 0.004 | 23260 ± 481 | 4029 ± 46 |
| Combination of all variabilities | 48.00 ± 0.55 | 23600 ± 2034 | 4100 ± 266 |
| Experiments | 48.056 ± 0.008 | 24600 ± 1300 | 4400 ± 1300 |

**2) Measurement uncertainty in PM magnetic characterization**

An expanded measurement uncertainty (95 % coverage interval) of 1.5 % is associated with the hysteresisgraph measurement procedure adopted in INRiM laboratories [51]. Therefore, a multiplicative correction coefficient for the PM magnetization characteristics was randomly extracted from a Gaussian statistical distribution having mean value equal to 1 and standard deviation equal 0.0075.

**3) Uncertainty in PM orientation**

The stochastic analysis applied an additional rotation angle on top of the 1.2° deterministic one. The value of this angle was extracted from a uniform distribution having bounds equal to 0.68° and 1.72°. These values are a consequence of the dimensional tolerances of sintered NdFeB PMs (±0.05 mm [52]) and the H10 mechanical tolerance recommended for CNC production of the PP rings [53]. The lower and upper bounds are obtained with the smallest pocket and largest PM and with largest pocket and smallest PM, respectively.

**4) Uncertainty in PM positioning**

In the previous paragraph, we accepted that the pockets could be up to 0.235 mm larger than the actual dimensions of the PMs to motivate a ±1.72° rotation of the PMs inside their pockets. Considering a dimension of the PM decreased by its 0.05 mm tolerance, the position of each PM can vary within a range of ±0.14 mm in the plane. These values were used as lower and upper limits of a uniform statistical distribution of the PM position deviation in the MC simulations.

**5) Combination of all stochastic variabilities**

Finally, an MC simulation was performed, accounting for all of the above sources of variability that were assumed to be uncorrelated.

## VI. DISCUSSIONS

The three deterministic causes of discrepancies between the homogeneity of the design outcomes and the prototype measurements have different impacts. They were studied in strict sequence, leading to variation in $DIS_1$ (resp. $DIS_2$) from the design values of 1720 ppm (316 ppm) to 6960 ppm (1340 ppm) (different PM characteristics), to 6520 ppm (1150 ppm) (ring positioning), and to 23200 ppm (4028 ppm) (max PM orientation due to torque). Corresponding relative variations are: +4.05 (+4.24), –0.94 (–0.86) and +3.56 (+3.50), respectively. The relative variations evidence the significant effect of both using different PM types in the array and the variation of PMs orientation due to mechanical forces. Conversely, the deviation of the ring positions along the *z*-axis with respect to their reference values, only causes relatively low effects.

This observation has a direct consequence for magnet designs. For example, the adoption of different PM types should be avoided if possible. Even when the PMs are of the same type (e.g. N52 in our example), their magnetic characteristics could slightly differ, introducing a potential cause of variability between the ideal prediction and the final realization.

The modification of PM orientation due to forces could be properly considered during the magnet optimization, either by reducing the likelihood of magnet rotation or by milling the magnet pockets at a slight angle in the opposite direction to compensate for it. This result also directly affects the choice of manufacturing tolerances, particularly in ring production.

Rather than providing strategies for compensation during the design phase, the analysis of potential stochastic effects allowed us to assign an uncertainty value to the design prediction.

The most significant stochastic effect was attributed to the measurement uncertainty in PM magnetic characterization. This effect introduces a bias in the expected value of $DIS_1$ of +0.9 % (23400 ppm instead of 23200 ppm) with a relative expanded uncertainty of ±7.3 % (95 % coverage interval). At the same time, a bias in the expected value of $DIS_2$ is +1.4 % (4080 ppm instead of 4028 ppm) with a relative expanded uncertainty of ±5.4 %. The sensitivity of the expanded uncertainty of $DIS_1$ (resp. $DIS_2$) versus the expanded uncertainty in PM characterization is 115 $10^3$ ppm/T (resp. 15 $10^3$ ppm/T).

The uncertainty in the actual orientation of the PMs, caused by the mechanical tolerances of the structure, introduces a smaller bias effect in the expected value of $DIS_1$ (23370 ppm against 23200 ppm, that is +0.7 %) with an expanded uncertainty of ±3.2 % (with a 95 % coverage interval). The bias in the expected value of $DIS_2$ is +0.1 % (4032 ppm instead of 4028 ppm) with an expanded uncertainty of ±2.2 %. The sensitivity of the expanded uncertainty of $DIS_1$ (resp. $DIS_2$) versus the maximum deviation of angle orientation is 83.4 $10^3$ ppm/rad (resp. 9.92 $10^3$ ppm/rad).

Weaker bias effects are found on the expected value of $DIS_1$ and $DIS_2$ due to variability in the PM samples (around +0.4 %) and PM positioning (less than 0.3 %). The sensitivity of the expanded uncertainty of $DIS_1$ (resp. $DIS_2$) versus the expanded uncertainty in PM characterization is 54 $10^3$ ppm/T (resp. 5.6 $10^3$ ppm/T), while the sensitivity of the expanded uncertainty of $DIS_1$ (resp. $DIS_2$) versus the maximum deviation of PM positioning is 3.4 $10^6$ ppm/m (resp. 0.33 $10^6$ ppm/m).

The combination of all the stochastic effects causes an increase of the expected $DIS_1$ value of +1.7 % (23600 ppm instead of 23200 ppm) with an expanded uncertainty of ±8.6 % (95 % coverage interval). The bias in the expected value of $DIS_2$ is +1.7 % (4100 ppm instead of 4028 ppm) with an expanded uncertainty of ±6.5 %.

A good agreement is found measurement and corrected modelling results, both for the $B_0$ field value and its homogeneity in the DSV. The 95 % coverage interval of the measured $B_0$ field average is (48.048 ÷ 48.064) mT while the model gives (47.45 ÷ 48.55) mT. For $DIS_1$ the coverage intervals for the experiment and the model are (23300 ÷ 25900) ppm and (21600 ÷ 25600) ppm, respectively. For $DIS_2$ the coverage intervals for the experiment and the model are (3100 ÷ 5700) ppm and (3800 ÷ 4400) ppm, respectively. A superposition of the coverage intervals is always found, demonstrating consistency between measurements and modelling predictions.

It should be noted that this study did not analyze the effects of the magnetic field's temporal stability. This is also crucial for practical applications involving long-term data acquisition, as it ensures the reliability of the data, and will require specific detailed analysis.

## VII. Conclusions

This study used a reference magnet array design to investigate the main factors responsible for discrepancies between the expected and measured performances.

Understanding how choices made during magnet construction impact the magnet performance, potentially helps in achieving more effective and reliable designs. Furthermore, analyzing sensitivity to stochastic effects of variability highlights the importance of optimizing the magnet using a configuration that is more robust against variations in the input data. All of this has a direct effect on the performance of the scanner in terms of image quality.

Finally, the results of the analysis can be leveraged to select the manufacturing tolerances in a more informed manner and, consequently, eliminate or at least simplify the shimming phase. In the latter case, the results may pave the way for an 'informed shimming', whereby the variability in the positioning and orientation of additional magnets can be substantially reduced.

In both cases, reducing production costs is feasible and would be relevant and useful to the wider portable MRI community.

## Datasets

All datasets are available on the Zenodo repository https://doi.org/10.5281/zenodo.15879992.

## Acknowledgments

The authors thank Dr. Lukas Winter and Dr. Martin Häuer from PTB for having provided the mechanical frame of the magnet. The authors also thank Dr. Tom O'Reilly for sharing all the data necessary for simulating the magnetic field distribution generated by the array.

The study was performed in the framework of 22HLT02 A4IM project. The project 22HLT02 A4IM has received funding from the European Partnership on Metrology, co-financed by the European Union's Horizon Europe Research and Innovation Programme and by the Participating States.

## Appendix A

The measurement models of parameters $DIS_1$ and $DIS_2$ are:

$$DIS_1 = \frac{B_{x,max} - B_{x,min}}{B_{x,mean}} \cdot C_{off,1} \quad \text{(A.1)}$$

$$DIS_2 = \sqrt{\frac{1}{N}\sum_i \left(B_{x,i} - B_{x,mean}\right)^2} \cdot C_{off,2} \quad \text{(A.2)}$$

where:

- $B_x$ is the $x$-component of the magnetic field $B$. The subscripts *max*, *min* and *mean* indicate the maximum, minimum and the mean measured values. Its uncertainty value is computed below.
- $C_{off,1}$ and $C_{off,2}$ are two unitary coefficients whose uncertainty takes into account the influence of the magnet array position with respect to the origin of the measurement system. The array position in the measurement system was uncertain up to ±2 mm along each Cartesian axis with a uniform distribution. Simulations showed that moving the DSV according to these limits causes the values of the $DIS_1$ and $DIS_2$ to change by a maximum of 4.8 % and 1 %, respectively. This resulted in a standard relative uncertainty equal to 14000 ppm and 3000 ppm for the $DIS_1$ and $DIS_2$ values, respectively.
- $N$ is the total number of measured values and equals to 4224.

The measurement model for the $x$-component of the magnetic flux density, $B_x$, is:

$$B_x = B_{x,meas} + C_{noise} + C_{res} + C_{avg} + C_{pos} + C_{temp} \quad \text{(A.3)}$$

where:

- $B_{x,meas}$ is the measured value of the $x$-component of the magnetic flux density in a specific spatial point inside the DSV. Its relative uncertainty was determined by calibration with a reference NRM probe and equals 20 ppm;
- $C_{noise}$ is a zero average coefficient whose uncertainty accounts for the measurement noise;
- $C_{res}$ is a zero average coefficient whose uncertainty accounts for the instrument digital resolution;
- $C_{avg}$ is a zero average coefficient whose uncertainty accounts for the dimension of the sensor. Its uncertainty was obtained from the maximum sensor linear dimension (100 µm) multiplied by the maximum $B_x$ gradient within the DSV and assessed through simulations (13.56 mT/m);

TABLE A

COLLECTION OF ALL THE UNCERTAINTY CONTRIBUTIONS OF THE $B_X$, $DIS_1$ AND $DIS_2$ MEASURED VALUES. $u(x)$ AND $U(x)$ ARE THE STANDARD AND EXPANDED UNCERTAINTY VALUES, RESPECTIVELY. $K$ IS THE COVERAGE FACTOR AND PDF IS THE PROBABILITY DENSITY FUNCTION. THE UNCERTAINTY CONTRIBUTION WERE LINEARLY PROPAGATED TO THE $DIS_1$ AND $DIS_2$ VALUES AND ARE REPORTED IN THE LAST TWO ROWS.

| | Best Estimate | $u(x)$ | $K$ | $U(x)$ | PDF |
|---|---|---|---|---|---|
| $B_{x,\text{meas}}$ | $B_{x,\text{meas,i}}$ | $50\ 10^{-6}\cdot B_{x,\text{meas,i}}$ | 2 | $100\ 10^{-6}\cdot B_{x,\text{meas,i}}$ | Normal |
| $C_{\text{noise}}$ | 0 | < 1 $\mu$T | 2 | < 1 $\mu$T | Normal |
| $C_{\text{res}}$ | 0 | < 1 $\mu$T | 0.95√3 | < 1 $\mu$T | Uniform |
| $C_{\text{avg}}$ | 0 | < 1 $\mu$T | 0.95√3 | < 1 $\mu$T | Uniform |
| $C_{\text{pos}}$ | 0 | 3.9 $\mu$T | 0.95√3 | 6.4 $\mu$T | Uniform |
| $C_{\text{temp}}$ | 0 | $406\ 10^{-6}\cdot B_{x,\text{meas,i}}$ | 0.95√3 | $668\ 10^{-6}\cdot B_{x,\text{meas,i}}$ | Uniform |
| $C_{\text{off,1}}$ | 1 | $0.014\cdot DIS_1$ | 0.95√3 | $0.023\cdot DIS_1$ | Uniform |
| $C_{\text{off,2}}$ | 1 | $0.003\cdot DIS_2$ | 0.95√3 | $0.005\cdot DIS_2$ | Uniform |
| | | | | | |
| $DIS_1$ | 24638 ppm | 642 ppm | 2 | 1283 ppm | Normal |
| $DIS_2$ | 4353 ppm | 643 ppm | 2 | 1286 ppm | Normal |

- $C_{\text{pos}}$ is a zero average coefficient whose uncertainty accounts for the uncertain position of the magnetic field probe both due to the automatic movement system and the mechanical tolerance of the 3D printed holder. The centering of the magnet array within the area scanned by the movement system has been measured with a resolution of ±0.5 mm and the tolerance of the probe holder is ±0.5 mm as well. By combining these contributions, the $C_{\text{pos}}$ uncertainty was obtained by multiplication with the maximum $B_x$ gradient within the DSV assessed through simulations (13.56 mT/m);
- $C_{\text{temp}}$ is a zero average coefficient whose uncertainty accounts for the temperature variation during the measurements. The measurement temperature equals (23.7±0.32) °C. Expressions (1.a) and (1.b) allowed to compute the uncertainty associated with the coefficient $C_{\text{temp}}$.

Table A collects all the uncertainty contributions together with their standard and expanded uncertainty values. The last two rows of the table report the uncertainty values linearly propagated to the $DIS_1$ and $DIS_2$ measured values. All contributions have been propagated in accordance with [55].

APPENDIX B

To extend the dipole modeling approach to account for magnet interactions and nonlinearities, the 'working point' of each magnet has to be computed according to its linear or nonlinear magnetic characteristics. The total magnetic field inside magnet $i$ ($H_v^{(i)}$) is given by the sum of the 'internal' magnetic field ($H_{v,\text{int}}^{(i)}$), represented by the demagnetizing factor $\aleph$,

$$H_{v,\text{int}} = -\frac{\aleph}{\mu_0}\cdot J_v \tag{B1}$$

and the superposition of the 'external' contributions ($H_{v,\text{ext}}$) coming from all other magnets, evaluated in the barycenter of magnet $i$ by using Eqns. (5). The demagnetizing factor is given by

$$\aleph = \frac{L_u L_w}{L_u L_v + L_u L_w + L_v L_w} \tag{B2}$$

Therefore, for the generic magnet $i$, we can write the total magnetic field as:

$$H_v^{(i)} = -\frac{\aleph^{(i)}}{\mu_0} J_v^{(i)} + \left(\sum_{p\neq i}^{P} J_v^{(p)}\, \boldsymbol{F}_{p,i}\right)\cdot \boldsymbol{v}^{(i)} \tag{B3}$$

where $\boldsymbol{v}^{(i)}$ is the unit vector of the local axis $v$ of magnet $i$, $\boldsymbol{F}_{p,i}$ is the coefficient of interaction between magnet $i$ and the generic magnet $p$, being $P$ the total number of magnets in the array. Note that, in (B3), the field contribution of the generic magnet $p$ is projected along the axis of magnetization $\boldsymbol{v}^{(i)}$ of magnet $i$. Introducing the magnetic characteristic, $J_v^{(i)}$ is expressed as

$$J_v^{(i)} = \mu_0 \mu_M H_v^{(i)} + J_r^{(i)} \quad \text{(linear } H\text{-}J \text{ characteristic)} \tag{B4a}$$

$$J_v^{(i)} = g\left(H_v^{(i)}\right) + J_r^{(i)} \quad \text{(nonlinear } H\text{-}J \text{ characteristic)} \tag{B4b}$$

where $g$ is the monotonic function representing the $H$-$J$ magnetic characteristic.

Making reference to the nonlinear case, Eqn. (B4b) can be solved using the Fixed Point (FP) technique [54], introducing an arbitrary linear term (slope $\mu_{\text{FP}}$) plus a residual $R$ to be iteratively computed:

$$J_v^{(i)} = g\left(H_v^{(i)}\right) + J_r^{(i)} = \mu_{\text{FP}} H_v^{(i)} + R^{(i)} + J_r^{(i)} \tag{B5}$$

By elaborating on (B3) and considering (B5), we can derive the final equation for the generic magnet $i$:

$$-\aleph^{(i)}\mu_{\text{FP}} H_v^{(i)} - \mu_0 H_v^{(i)} + \mu_0 \mu_{\text{FP}} \sum_{p\neq i}^{P} \left(\boldsymbol{F}_{p,i}\cdot \boldsymbol{v}^{(i)}\right) H_v^{(p)} = \aleph^{(i)} J_r^{(i)} - \mu_0 \sum_{p\neq i}^{P} \left(\boldsymbol{F}_{p,i}\cdot \boldsymbol{v}^{(i)}\right) J_r^{(p)} + \aleph^{(i)} R^{(i)} - \mu_0 \sum_{p\neq i}^{P} \left(\boldsymbol{F}_{p,i}\cdot \boldsymbol{v}^{(i)}\right) R^{(p)} \tag{B6}$$

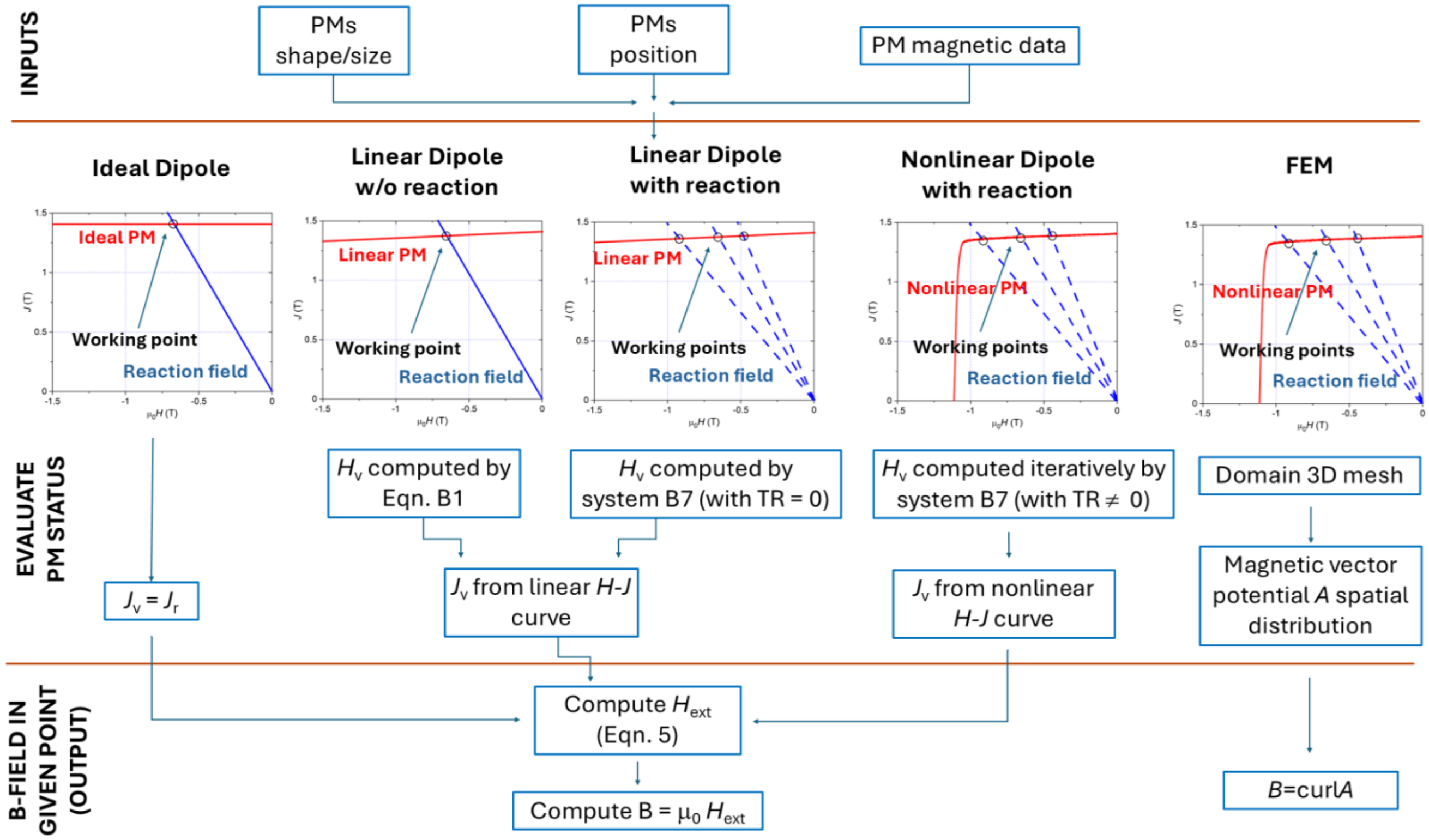

**Fig. B1.** Schematic diagram showing the different ways of modelling the PMs array.

which simplifies to the linear model when $\mu_{FP} = \mu_0 \cdot \mu_M$ and $R = 0$.

By writing (B6) for all magnets ($i = 1,\ldots,P$), we obtain the final system of equations:

$$\boldsymbol{M}\,\boldsymbol{H}_v = \boldsymbol{T}_J + \boldsymbol{T}_R \tag{B7}$$

where $\boldsymbol{H}_v$ is the vector containing the unknown values of the main components (local $v$-axis) of the magnetic field inside each magnet, $\boldsymbol{M}$ is a matrix accounting for the PM shape and interaction between PMs, and $\boldsymbol{T}_J$, $\boldsymbol{T}_R$ are known terms including magnetisation and FP residuals, respectively; the latter being zero for linear magnetic characteristics. The system of equations (B7) can be solved iteratively, starting from an arbitrary initial value of $\boldsymbol{T}_R$ (usually zero), until a given convergence tolerance is reached. At each FP iteration, the residual values $R^{(i)}$ are updated as follows:

$$R^{(i)} = g\left(H_v^{(i)}\right) - \mu_{FP} H_v^{(i)} \tag{B8}$$

Usually, some tens of iterations (< 50) are sufficient to reach a relative convergence tolerance lower than $10^{-7}$.

Finally, in Figure B.1 a simple diagram schematically showing the different ways of modelling the PMs array is shown.

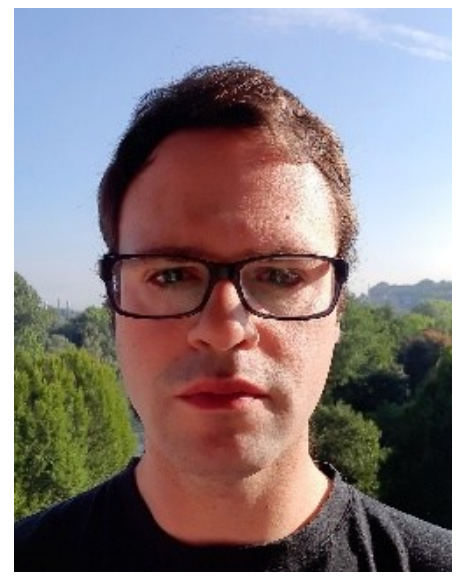

**Umberto Zanovello** was born in Torino, Italy, in 1988. He received the B.S. and M.S. degrees in Electrical Engineering and the Ph.D. degree in Electrical, Electronics and Communications Engineering from Politecnico di Torino, in 2011, 2014, and 2019, respectively. He has been a permanent researcher with the Istituto Nazionale di Ricerca Metrologica (INRiM), Torino, since 2021. His research interests include MRI safety and hardware design, with a particular focus on Low-Field MRI and Radiofrequency coils.

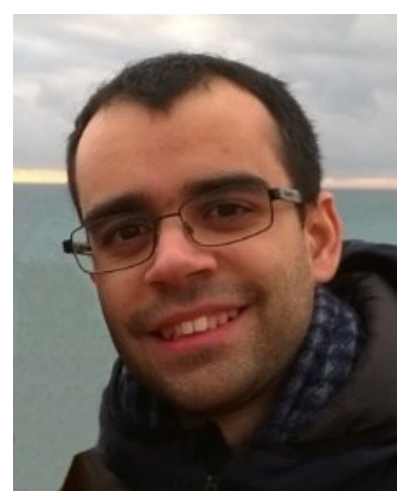

**Alessandro Arduino** received the B.Sc. degree in mathematics for engineering sciences, the M.Sc. degree in mathematical engineering, and the Ph.D. degree in electrical, electronics, and communications engineering from the Politecnico di Torino, Turin, Italy, in 2012, 2014, and 2018, respectively. Since 2018, he has been with the Istituto Nazionale di Ricerca Metrologica, Turin, Italy. He is the Maintainer of EPTlib, an open-source C++ library of magnetic resonance-based electric properties tomography methods. His research interests include mathematical and numerical modeling of electromagnetism applied to biomedical engineering and inverse problems.

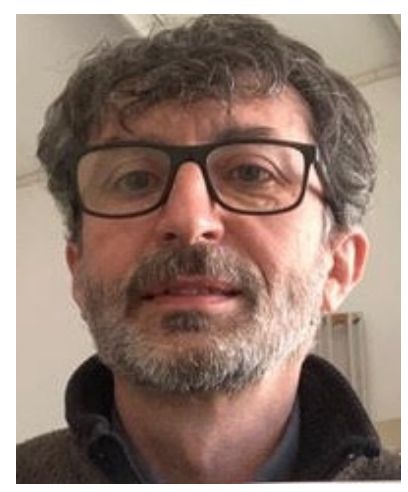

**Vittorio Basso** is an INRIM research scientist since 1995. His research interests are the physics of magnetic materials and spintronics. He is co-author of 160 peer reviewed publications. He is responsible of the INRIM magnetic measurement laboratory.

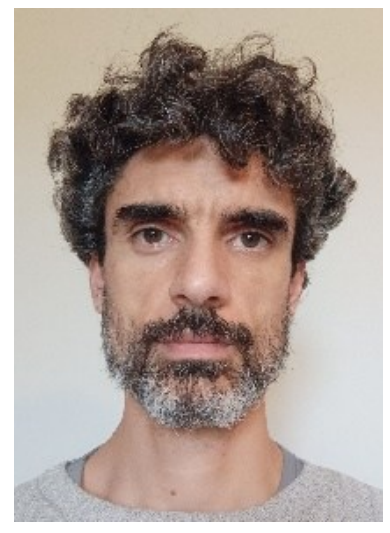

**Luca Zilberti** was born in Vercelli, Italy, in 1982. He received the B.Sc., M.Sc., and Ph.D. degrees in electrical engineering from Politecnico di Torino, Turin, Italy, in 2004, 2006, and 2010, respectively. In 2010 he joined the Istituto Nazionale di Ricerca Metrologica, Turin, where he is currently a senior researcher. In 2022 he got the national qualification as full professor for Electrical Engineering. His research interests include the development of computational methods for electromagnetic dosimetry and biomedical applications of electromagnetic fields. He currently serves in the governing committee of the study group on MR Safety of the International Society for Magnetic Resonance in Medicine (ISMRM). He coordinated the European EMPIR Project 18HLT05 "Quantitative MR-based imaging of physical biomarkers" (2019-2022) and currently coordinates the European EPM Project 24DIT01 APULEIO "Trustworthy and quality-assured quantitative magnetic resonance imaging" (2025-2028). In 2016 Dr. Zilberti was the recipient of the Young Scientist Award from the International Commission on Non-Ionizing Radiation Protection (ICNIRP).

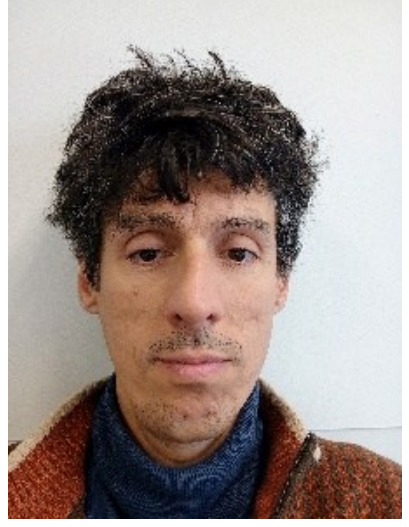

**Alessandro Sola** was born in Turin, Italy in 1984. He received the B.Sc. and the M.Sc. in physics from Università di Torino and Ph.D. degrees in physics from Politecnico di Torino, Turin, Italy in 2014. In the same year he joined the Istituto Nazionale di Ricerca Metrologica (INRIM) as a post-doc researcher and since 2020 as a permanent researcher. He leads the laboratory for the characterization of hard magnetic materials at INRIM and his research interest includes magnetic materials, spincaloritronics, and transverse thermoelectric effect.

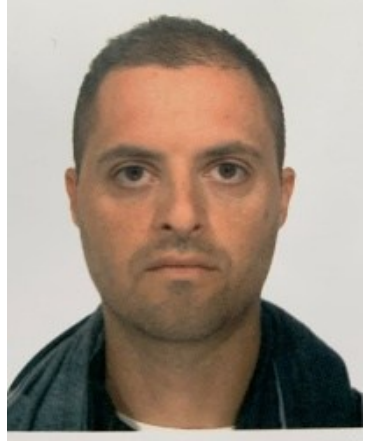

**Andrea Agosto** was born in Carmagnola, Italy, in 1981 and he received the high school diploma of electronics and telecommunications in 2000. He has been a technical collaborator and software developer at INRiM since 2001 and he is currently the Laboratory Manager and Activity Manager for calibration of AC magnetic sources and meters. He has been and continues to be Quality Assistant for Advanced Materials Metrology and Life Science Division and responsible for Quality implementation and maintenance/improvement of the INRiM QMS for the AC magnetic measurement laboratories of the ML Division. He co-drafts and co-revises technical procedures for performing calibrations, tests, and measurements.

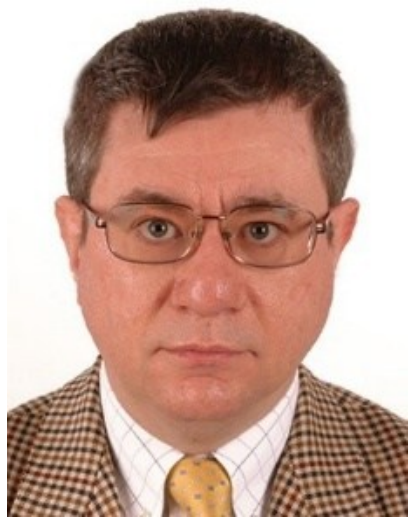

**Luca Toso** was born in Torino, Italy, in 1973 and he received the high school of electronics in 1992. He has been a technical collaborator at INRIM since 1994 and he is currently the Laboratory Manager and Activity Manager for calibration of analog/digital instruments for measuring static magnetic quantities. He has been and continues to be responsible for quality implementation and maintenance/improvement of the INRIM QMS for the DC magnetic measurement laboratories of the ML Division. He co-drafts and co-revises technical procedures for performing calibrations, tests, and measurements.

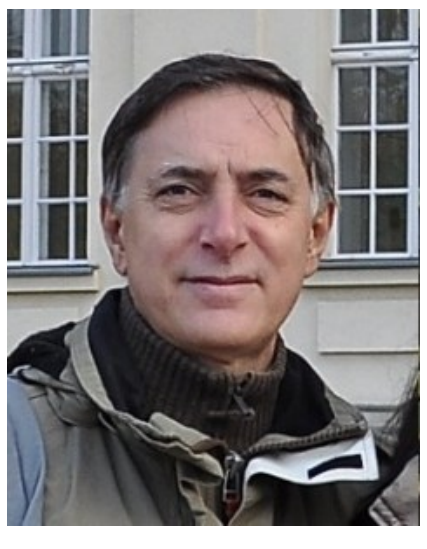

**Oriano Bottauscio** was born in Turin, Italy, in 1961. He received the M.Sc. degree in electrical engineering from Politecnico di Torino, in 1985. In 1987, he joined the Istituto Nazionale di Ricerca Metrologica (INRiM), formerly Istituto Elettrotecnico Nazionale

Galileo Ferraris, and got the level of the Research Director, in 2012. In the same year, he received the qualification as a Full Professor of electrotechnics. In 2015, he was a Member of the Board of Professor of the Ph.D. course in metrology with Politecnico di Torino. Since 2018, he has been in the role of Deputy Coordinator. He is the author of more than 250 journal articles, with around 2700 citations (H-index 27). His research interests include computational electromagnetism for biomedical applications, with a main emphasis on MRI safety and medical devices. He is the INRiM contact person in the EURAMET Technical Committee of Interdisciplinary Metrology.